\documentclass[11pt,a4paper]{article}

\usepackage[margin=1in]{geometry}
\usepackage{authblk} 
\usepackage{lineno}
\usepackage{times}
\usepackage{ulem}
\usepackage{amsmath,amssymb,amsfonts}
\usepackage{physics}
\usepackage{siunitx}
\usepackage{dsfont}
\usepackage{bbm}
\usepackage{mhchem}
\usepackage{makecell}
\usepackage{graphicx}
\usepackage{float}
\usepackage{subfigure}
\usepackage{booktabs}
\usepackage{array}
\usepackage{diagbox}
\usepackage{multirow}
\usepackage{tabularx}
\usepackage{xcolor}
\definecolor{darkblue}{HTML}{004D6B}
\definecolor{darkred}{HTML}{8c1515}
\definecolor{darkgreen}{HTML}{006400}
\usepackage{hyperref}
\hypersetup{
    pdftitle={Robust Teleportation of a Surface Code},
    colorlinks=true,
    linkcolor=darkred,
    citecolor=darkblue,
    urlcolor=darkred
}

\usepackage{titlesec}

\titleformat{\section}
  {\normalfont\Large\bfseries}
  {\thesection}
  {1em}
  {}

\titleformat{\subsection}
  {\normalfont\large\bfseries}
  {\thesubsection}
  {1em}
  {}

\titleformat{\subsubsection}
  {\normalfont\normalsize\bfseries}
  {\thesubsubsection}
  {1em}
  {}

\begin{document}

\begin{center}
{\LARGE\bfseries
Physics-informed Bayesian Optimization for Quantitative High-Resolution Transmission Electron Microscopy
\par}

\vspace{1em}

\textbf{Xiankang Tang}$^{1}$,
\textbf{Yixuan Zhang}$^{1}$,
\textbf{Juri Barthel}$^{2}$,
\textbf{Chun-Lin Jia}$^{3}$,
\textbf{Rafal E. Dunin-Borkowski}$^{2}$,
\textbf{Hongbin Zhang}$^{1,\dagger}$,
\textbf{Lei Jin}$^{2,\dagger}$

\vspace{0.8em}

$^{1}$ Institute of Materials Science, Technical University of Darmstadt,  
64287 Darmstadt, Germany

$^{2}$ Ernst Ruska-Centre for Microscopy and Spectroscopy with Electrons,  
Forschungszentrum Jülich GmbH, 52425 Jülich, Germany

$^{3}$ School of Microelectronics, Xi’an Jiaotong University, Xi’an 710049, China

\vspace{0.8em}

$^{\dagger}$ Corresponding authors:  

\href{mailto:hongbin.zhang@tu-darmstadt.de}{hongbin.zhang@tu-darmstadt.de},
\href{mailto:l.jin@fz-juelich.de}{l.jin@fz-juelich.de}

\vspace{0.8em}
\today
\end{center}

\vspace{2em}

\begin{abstract}

Quantitative high-resolution transmission electron microscopy (HRTEM) provides an indispensable means to understand the structure-property relationships of a material in atomic dimensions. Successful quantification requires reliable retrieval of essential atomic structural information despite artifacts arising from unwanted but practically unavoidable imaging imperfections. Experimental observation carried out in tandem with model-based iterative image simulation shows vast applications in quantitative structural and chemical determination of objects spanning zero to three dimensions~\cite{urbanProgressAtomicresolutionAberration2023}. However, the large number of parameters involved in the simulations makes the current multi-step, user-supervised iterative approach highly time-consuming, thereby restricting its application primarily to small sample areas and to experienced users. In this work, we implement and apply a physics-informed Bayesian optimization (BO) framework to advance HRTEM quantification towards full automation and large-field-of-view analysis. Unlike conventional optimization approaches, our method adopts a stepwise strategy that fully leverages the strength of BO in handling high-dimensional parameters, while its probabilistic engine rigorously and efficiently refines the parameter space to enable rapid quantification. Using a BaTiO$_3$ single crystal that contains heavy, medium and light elements as a model system, we demonstrate that the three-dimensional crystal structure can be determined from a single HRTEM image with a three to four order-of-magnitude improvement in time efficiency. This approach thus opens new avenues for fast and automated image quantification over larger sample volumes and promisingly, for \textit{in situ} applications.

\noindent\textbf{Keywords:} quantitative high-resolution transmission electron microscopy, Bayesian optimization, three-dimensional atomic structure, BaTiO$_3$

\end{abstract}


\section{Introduction}

High-resolution (HR) transmission electron microscopy (TEM) enables direct atomic resolution observation of crystal lattices, defects, surfaces, and interfaces, whose structures are often decisive for the microscopic properties of materials~\cite{urbanProgressAtomicresolutionAberration2023}. Under \textit{in situ/operando} conditions, HRTEM also provides critical insights into phase transformations, lattice deformations, material synthesis, and catalysis processes, etc., whichhaves significantly advanced the field of materials science (\textit{e.g.}, \cite{sunInSituTransmissionElectron2023,liReviewRecentProgress2022,zhouSituElectronMicroscopy2025,rauretAdvancesSituOperando2025,chaoSituEmergingTransmission2023,xingAtomicscaleOperandoObservation2022,jeongSubsurfaceOxygenVacancy2025,weiUnconventionalTransientPhase2020,weiSituObservationPointDefectInduced2021}). For example, recent \textit{in situ} atomic-resolution TEM studies have revealed the surface reconstruction and depolarization mediated by subsurface oxygen vacancies in ferroelectric BaTiO$_3$ (BTO)~\cite{jeongSubsurfaceOxygenVacancy2025}.

The essential structural information contained in HRTEM images is often distorted by unwanted but practically unavoidable effects resulting from imaging imperfections (such as residual aberrations) as well as subtle orientation misalignment of samples. It becomes further blurred by effects limiting spatial resolution, such as partial coherence of the electron beam and noise (in terms of image spread $\sigma$ \cite{jiaAtomicScaleMeasurementStructure2013}\cite{jiaDetermination3DShape2014}\cite{jinAtomicResolutionImaging2017} with a major contribution of magnetic Johnson noise~\cite{uhlemannThermalMagneticField2013a}), and by imperfections of the recording device described by the modulation transfer function (MTF) \cite{thustHighResolutionTransmissionElectron2009}. In addition, a linear relationship between the observed peak intensity and the actual atomic column configuration is not always guaranteed due to the nonlinear nature of electron phase-contrast imaging~\cite{jiaQuantitativeHRTEMIts2018}. Successful retrieval of this essential information therefore requires quantitative evaluation of experimental HRTEM images, which has been achieved previously through image reconstruction of focal series~\cite{jiaAtomicStructureSigma31999,jiaInvestigationAtomicDisplacements1999,jiaAtomicScaleAnalysisOxygen2005,houbenAtomicprecisionDeterminationReconstruction2006} or electron holography ~\cite{lehmannTutorialOffAxisElectron2002,winklerAbsoluteScaleQuantitative2018a}, and model-based image simulation~\cite{jiaAtomicScaleMeasurementStructure2013,jiaDetermination3DShape2014,jinAtomicResolutionImaging2017,geAtomicScaleObservationOffCentering2022,jinUnderstandingStructuralIncorporation2022}. Compared with reconstruction methods that recover the electron wave function at the specimen exit plane, image simulations relate the image information to atomic structure data, such as the configuration of individual atomic columns and their relative positions.

\begin{figure}[htbp] 
   \centering
   \includegraphics[width=0.95\columnwidth]{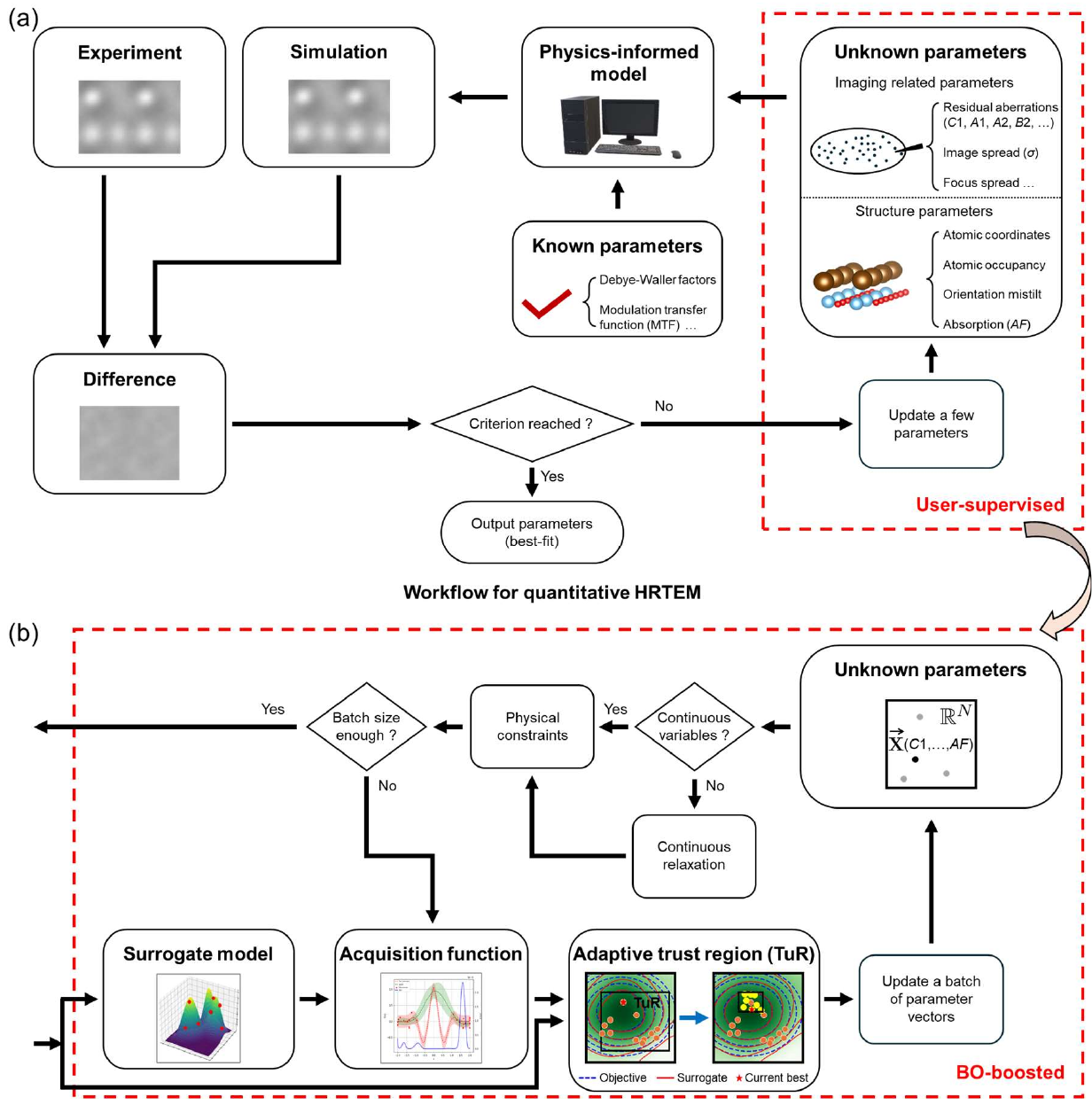} 
   \caption{{\bf Conventional and proposed Bayesian optimization boosted workflow for quantitative HRTEM}. (a) Conventional workflow based on user-supervised, multi-step iteration. The experimental image is set as the input. The unknown parameters are initialized in the first iteration and iteratively updated until the best fit between simulation and experiment is achieved. (b) Implemented BO workflow replacing the user-supervised procedure in (a). In BO, a batch of input parameter vectors $\mathbf{X} \in \mathbb{R}^n$ is initialized for subsequent optimization. Respective difference images between experiment and simulation are used to train a surrogate model and to dynamically adjust the TuR in $\mathbb{R}^n$, within which a new search is conducted using a batch of parameter vectors proposed by the acquisition function. Continuous relaxation is introduced to handle both continuous and discrete parameters and physical constraints are also enforced before generating new candidate $\mathbf{X}$.}
   \label{fig: workflow}
\end{figure}

A workflow for the model-based quantitative HRTEM has been demonstrated (\textit{e.g.}~\cite{jiaQuantitativeHRTEMIts2018, mobusStructureDeterminationMetalceramic1994, mobusRetrievalCrystalDefect1996}) and here an extended version is illustrated in Fig.~\ref{fig: workflow}(a). In the workflow, simulations are generated using a physics-informed model in which all relevant parameters are explicitly specified. These include known quantities obtained from independent measurements (such as MTF, types of atoms, Debye-Waller factors, \textit{etc}.), as well as unknown parameters to be determined—belonging to the unknown parameter space, denoted as $\mathbb{R}^n$. Simulated images are compared with the experimental image according to a user-defined criterion. Widely used criteria include image contrast \cite{jiaAtomicScaleMeasurementStructure2013}\cite{jiaDetermination3DShape2014}\cite{geAtomicScaleObservationOffCentering2022}, defined as the standard deviation (std) of the intensity distribution of an image when its mean value is normalized to unity; image similarity, evaluated by the normalized cross-correlation coefficient (\textit{NCC})~\cite{mobusStructureDeterminationMetalceramic1994,mobusRetrievalCrystalDefect1996,thustQuantitativeHighspeedMatching1992} or the structural similarity index metric \cite{edererImageDifferenceMetrics2022,wangStudyAberrationDetermination2025,zhouwangImageQualityAssessment2004}; and goodness of fit, characterized by the $\chi^2$~\cite{kingDeterminationThicknessDefocus1993a} statistic or the R-factor~\cite{smithCalculationDisplayComparison1982a} as employed in X-ray and neutron refinement. The parameter space is then explored iteratively until the best match between simulation and experiment is achieved.

The unknown parameters can be generally grouped into two categories: imaging related parameters (including residual aberrations of the imaging lens system, image spread $\sigma$, \textit{etc}.) and structure parameters (\textit{e.g.}, atomic positions, atomic occupancies, electron absorption factor $\textit{AF}$~\cite{jiaAtomicScaleMeasurementStructure2013}\cite{jiaDetermination3DShape2014}\cite{jinAtomicResolutionImaging2017}). Here, the notation for aberrations follows Ref.~\cite{uhlemannResidualWaveAberrations1998}, including defocus $C1$, twofold astigmatism $A1$, threefold astigmatism $A2$, axial coma $B2$, spherical aberration $C3$, fourfold astigmatism $A3$, star aberration $S3$, and so forth. Absorption describes the loss of electrons throughout the imaging process, \textit{i.e.}, electrons that are not registered by the camera because of multiple origins, such as electron back-scattering and blockage by apertures~\cite{jiaDetermination3DShape2014}. Although this factor is related to imaging in principle, it is listed here as a structure parameter due to the way we account for it in simulations. As an approximate treatment in image simulations, an imaginary part is introduced to the atomic scattering potential (\textit{i.e.}, absorptive potential), adopting the approach of Hashimoto, Howie and Whelan~\cite{hashimotoAnomalousElectronAbsorption1962}. It is apparent, for example, from Fig.~\ref{fig: workflow} (a), that the unknown parameter space to be determined is hyperdimensional. In particular, the structure parameters increase dramatically as the region to be studied increases, for instance, in the case of determination of the three-dimensional (3D) shape of nanoscale crystals ~\cite{jiaDetermination3DShape2014}. The current approach, based on multi-step, user-supervised iteration, usually optimizes the model by changing the values of a few parameters at each iteration. It is thus a tedious and very time-consuming procedure to optimize all the parameter values collaboratively, even for experienced users. For extending the application range of model-based image quantification to larger sample volumes and, ultimately, to structural variations in the time domain, optimization efficiency represents a central bottleneck. While previous \textit{in situ} atomic-resolution TEM studies have shown that transient or intermediate states can be captured by deliberately slowing down transition kinetics using external stimuli (\textit{e.g.}, electron-beam irradiation or thermal activation~\cite{xingAtomicscaleOperandoObservation2022,jeongSubsurfaceOxygenVacancy2025,weiUnconventionalTransientPhase2020,weiSituObservationPointDefectInduced2021}), the practical limitation in such studies often lies not in data acquisition, but in the time-consuming post-acquisition image quantification.

Recent advances in machine learning (ML) have shed new light on the automated experimentation and advanced post-acquisition data analysis in TEM, stimulating a paradigm shift towards real-time analysis and closed-loop microscope operation (\textit{e.g.}, \cite{kalininMachineLearningScanning2022b, kalininMachineLearningAutomated2023a, botifollMachineLearningElectron2022d, chengReviewSituTransmission2022a}). Extensive applications have been demonstrated using ML to perform image denoising, image labelling (\textit{e.g.}, classifying image region quality, determining crystal structure and orientations) and semantic segmentation (\textit{e.g.}, identifying atom species, positions and defects). A comprehensive overview of recent advances in this field can be found in Ref.~\cite{edeDeepLearningElectron2021}. In the context of phase retrieval, convolutional neural network (CNN) has been applied to 4D-STEM and HRTEM~\cite{friedrichPhaseObjectReconstruction2023,lethlarsenReconstructingExitWave2023}. In the meantime, several fundamental limitations of the CNN-based approaches should be noted. On the one hand, CNN typically requires huge and high-quality training datasets and often struggles with generalization across diverse experimental conditions; on the other hand, CNN-based algorithms inherently focus on learning pixel-level statistical correlations, while paying less attention on the underlying physical mechanisms. To address these challenges, physics-driven ML — a class of methods that incorporate physical prior knowledge into ML algorithms \cite{mengWhenPhysicsMeets2025} — offers a new perspective of mitigating the shortage of training data, increasing model generalizability and ensuring physical interpretability of results.

In this work, we implement a physics-informed Bayesian optimization (BO) approach to boost the image quantification of HRTEM, which allows the imaging related parameters and essential structure features to be determined with remarkably improved time efficiency as compared with conventional iterative methods. We begin with a detailed description of the proposed BO framework in the next section. This approach is then applied to experimentally characterized BTO, consisting of heavy, medium, and light elements, to determine the 3D crystal structure from a single HRTEM image and provide direct evidence of depolarization in size-limited BTO. It is demonstrated that our method effectively explores the high-dimensional parameter space and maintains promising adaptability across different material systems. Our work thus offers a new perspective on quantitative HRTEM analysis, paving the way for more efficient and automated experimentation. In future applications of the method, we aim to reconstruct 3D atomic structures over larger fields of view, providing a robust and efficient approach for generating digital twins of materials.

\section{Bayesian Optimization}

BO is a data-efficient optimization framework particularly suited for problems in which evaluating the objective function is costly, time-consuming, or experimentally demanding~\cite{frazierTutorialBayesianOptimization2018}. Instead of exhaustively sampling the parameter space, BO constructs a surrogate model—mostly based on Gaussian process (GP) regression—to approximate the underlying function including uncertainty~\cite{frazierTutorialBayesianOptimization2018, rasmussenGaussianProcessesMachine2006}. This uncertainty-aware modeling allows the algorithm to more efficiently select new candidates through an acquisition function~\cite{frazierTutorialBayesianOptimization2018}. As a result, fewer evaluations of the actual model are necessary, leading to a more rapid convergence to an optimal solution, even in noisy and high-dimensional settings. Owing to these advantages, BO has been widely adopted in materials science for tasks such as composition/phase optimization, autonomous materials discovery, and experimental data refinement/interpretation (cf.~\textit{e.g.}, \cite{xueAcceleratedSearchMaterials2016, yuanAcceleratedDiscoveryLarge2018, balachandranExperimentalSearchHightemperature2018, raoMachineLearningEnabled2022, liSequentialClosedloopBayesian2024, kusneOntheflyClosedloopMaterials2020,gayon-lombardoDeepKernelBayesian2025, zhangAutonomousAtomicHamiltonian2023a}). More recently, it has also been applied to automated aberration correction in scanning TEM~\cite{pattisonBEACONAutomatedAberration2025}, highlighting its versatility across experimental and computational workflows.

The workflow of BO is implemented (Fig.~\ref{fig: workflow}(b)) to replace the user-supervised parameter update step shown in Fig.~\ref{fig: workflow}(a). The objective function (criterion function) is defined as the mean squared error (MSE) below:

\begin{equation}
    \mathcal{L}_\mathrm{MSE}(\mathbf{X}) = \frac{1}{N} \sum_{i=1}^{N} \left( I_\mathrm{Sim}^{(i)} (\mathbf{X})- I_\mathrm{Exp}^{(i)} \right)^2,
\end{equation}
where $I_\mathrm{Sim}^{(i)}$ and $I_\mathrm{Exp}^{(i)}$ are intensities at pixel $i$ of the experimental and simulated images, respectively, $N$ is the number of pixels summed over and $\mathbf{X} \in \mathbb{R}^n$ denotes a vector of unknown parameter values (Fig.~\ref{fig: workflow}(b)). The same space is illustrated schematically in Fig.~\ref{fig: workflow} (b) in an alternative representation. A trust-region (TuR) BO (TuRBO)~\cite{erikssonScalableGlobalOptimization2020} strategy, which dynamically adapts the search domain within $\mathbb{R}^n$ during the optimization process, is employed to find the minimum of $\mathcal{L}_\mathrm{MSE}(\mathbf{X})$ (\textit{i.e.}, criterion in Fig.~\ref{fig: workflow}(a)). Unlike traditional BO, which often struggles with high-dimensional spaces and mixed discrete-continuous parameter types, our algorithm mitigates these limitations at the sampling stage by incorporating constraints—such as energy considerations for surface atoms—thereby ensuring, to a large extent, that the sampled points remain physically meaningful. To achieve computational efficiency in HRTEM quantification, we deliberately avoid algorithms that perform well on high-dimensional problems but incur prohibitively high computational costs, such as regional expected improvement for efficient TuR selection~\cite{namuraRegionalExpectedImprovement2024a} and feasibility-driven TuRBO~\cite{asciaFeasibilityDrivenTrustRegion2025a}.

As indicated in Fig.~\ref{fig: workflow}(b), a batch of input parameter vectors $\mathbf{X}$ and their corresponding $\mathcal{L}_\mathrm{MSE}(\mathbf{X})$ are used to train a surrogate model, \textit{i.e.}, via a GP~\cite{rasmussenGaussianProcessesMachine2006}. For this purpose, a covariance matrix is constructed using the Matern 2.5 kernel~\cite{gardnerGPyTorchBlackboxMatrixMatrix2021}, which provides a flexible trade-off between smoothness and modeling capacity. The code is implemented using the BoTorch library~\cite{balandatBoTorchFrameworkEfficient2020} and instantiated as a FixedNoiseGP~\cite{balandatBoTorchFrameworkEfficient2020}, \textit{i.e.}, a single-task exact GP with a prescribed observation noise level. The algorithm employs a relatively strong prior on kernel hyperparameters, which helps to mitigate overfitting and enhances generalization performance and numerical stability when the training data are normalized to have zero mean and unit variance~\cite{rasmussenGaussianProcessesMachine2006}.

After training the surrogate model, we use an acquisition functions based on parallel upper confidence bound (qUCB)~\cite{wilsonReparameterizationTrickAcquisition2017} to generate new candidates within the current TuR and adjust the hyperparameter $\beta$ to balance exploration and exploitation. Here, qUCB is defined as:

\begin{equation}\label{eq:qucb}
q\mathrm{UCB}(\mathbf{X}) = \mathbb{E}\left[\max\left(\mu + \left| \mathbf{\tilde{Y}} - \mu \right|\right)\right], \quad \mathbf{\tilde{Y}} \sim \mathcal{N}\left(\mu, \beta \cdot \frac{\pi}{2} \mathbf{\Sigma}\right)
\end{equation}

where $\mu$ is the posterior mean function of the surrogate model, representing the predicted values at the test points $\mathbf{X}$ based on the observed data, ${\mathbf{\Sigma}}$ is the corresponding posterior covariance function, and $\beta$ controls the scaling of the posterior covariance and thus modulates the degree of uncertainty injected into the samples $\tilde{Y}$. A larger $\beta$ encourages more exploratory behavior by increasing the variance of the samples, while a smaller $\beta$ favors exploitation by concentrating around the predictive mean. The TuR serves as a dynamic constraint in the search domain in $\mathbb{R}^n$: when the newly evaluated batch leads to a significant reduction in $\mathcal{L}_\mathrm{MSE}$, the region expands to encourage a broader exploration; conversely, if the improvement is insufficient, the region contracts to focus the search on more promising neighborhoods. Through this mechanism, qUCB can be optimized over a controllable and well-conditioned subset of the parameter space $\mathbb{R}^n$, thereby maintaining stable search performance in high-dimensional problems.

To reduce computational overhead and avoid maintaining multiple GPs, our implementation employs a single TuR. Although a TuR increases the risk of TuRBO becoming trapped in a local optimum, its performance is found to be satisfactory for the current task. If global exploration is required, the framework can be extended with a simple but effective warm-restart strategy ~\cite{martiMultistartMethodsCombinatorial2013,poloczekWarmStartingBayesian2016}. Once the TuR shrinks to the minimum size set, optimization can be restarted from a newly sampled set of initial points generated under different random seeds, while retaining all previously evaluated points in the training set and increasing $\beta$ to enhance the exploratory nature of qUCB. In practice, as the number of restarts increases, the probability of locating the global optimum increases \cite{leNonsmoothNonconvexStochastic2024}.

Acquisition functions in BO usually operate in a continuous space, whereas our optimization process inherently involves both continuous and discrete parameters. Such a mixed nature of parameters breaks the continuity assumption validated for gradient-based optimization of the acquisition function; thus, further treatment is required to maintain a continuous GP. A common strategy to address this issue is to perform continuous relaxation (CR) for the discrete parameters, for example, by representing categorical choices through probabilistic embeddings and optimizing the corresponding distributional parameters, as implemented in probabilistic reparameterization (PR) ~\cite{daultonBayesianOptimizationDiscrete2022}. However, PR requires Monte Carlo to calculate the expected value of the acquisition function in each iteration; it is resource-intensive and thus reduces computational efficiency, especially when the number of discrete data combinations is extremely large. Inspired by PR and Refs.~\cite{michaelContinuousRelaxationDiscrete2024,maddisonConcreteDistributionContinuous2017}, we introduce a CR-based method for the present workflow. We reformulate the optimization problem by introducing a discrete probability distribution $p(Z|\theta)$  parameterized by a vector of continuous parameters $\theta$  over a random discrete parameter $Z$ such as the number of atoms in an atomic column. Defining the vector $\mathbf{\Theta} = [\theta_1,\theta_2,\dots,\theta_n] \in \mathbb{R}^n$ as the set of continuous random variables, the number of atoms $\mathbf{Z} = [Z_1, Z_2,\dots, Z_m] \in \mathbb{Z}^m$ in each atomic column can be expressed as:
\begin{equation}
\mathbf{Z} = \lfloor \mathbf{\Theta} \rfloor + \mathbf{B}, \mathbf{B} \sim \rm{BERNOULLI}(\mathbf{\Theta}-\lfloor \mathbf{\Theta} \rfloor) 
\label{con:cr}
\end{equation}

With this operation, CR maps the discrete space to a continuous proxy via a Bernoulli distribution, which eventually allows the acquisition function to be treated as a continuous function. During the sampling process, randomness from a Bernoulli distribution is introduced to ensure that $\mathbf{Z}$ is an integer, while the soft representation of $\mathbf{\Theta}$ is preserved, and the bias caused by purely deterministic rounding, such as simply taking $\lfloor \mathbf{\Theta}\rfloor$ or $\lceil \mathbf{\Theta}\rceil$, is avoided. In comparison with PR, the present treatment avoids the use of Monte Carlo to compute the expectation of the acquisition function, thereby significantly improving computational efficiency. Meanwhile, it retains the capability for local exploration within the TuR while avoiding exhaustive global enumeration. Since the present treatment incorporates the distribution of $\mathbf{Z}$ solely at the sampling stage, it may cause divergence of the recommended result from the optimum. However, this divergence can be corrected by adopting the warm-start strategy described above. Given that each optimization run is inexpensive, the overall computational cost remains lower than that of Monte Carlo-based CR.

To improve reliability without sacrificing efficiency, we also incorporate constraints to exclude sampled points that violate known physical rules or pre-defined hypothesis, prior to entering the optimization loop. For instance, when optimizing atomic coordinates, each atom is assumed to remain within a physically admissible spatial range; configurations in which atoms violate pre-defined geometric constraints are regarded as energetically unfavorable and are therefore rejected. By filtering out such candidates in advance, the algorithm achieves substantially faster convergence while enhancing the physical interpretability of the resulting optimal configuration.

\section{Experimental Section}

To validate our approach by example, we select BTO that contains heavy, medium and light elements as a model system. An experimentally recorded HRTEM image along the [110] direction is used. The image was acquired under the negative spherical aberration imaging (NCSI) conditions~\cite{jiaAtomicResolutionImagingOxygen2003} on an FEI Titan 80-300 microscope, equipped with a field emission gun and an aberration correction system for the objective lens. Prior to image acquisition, the major aberrations were measured following the Zemlin tableau method~\cite{zemlinImageSynthesisElectron1978} and corrected to stay within the $\pi/4$ limit pertinent to the available spatial resolution. The available point resolution is better than 80 pm at an accelerating voltage of 300 kV. For image acquisition, a Gatan 2k$\times$2k UltraScan 1000P charge-coupled device camera was used with a sampling rate of 9.5 pm per pixel. The camera MTF was measured using the knife-edge method~\cite{thustHighResolutionTransmissionElectron2009} and input to the simulation as a known parameter.

To ensure quantitative analysis between the experimentally recorded and simulated images on the absolute contrast level, the recorded HRTEM image was normalized with respect to an image obtained subsequently from a nearby vacuum, and the resulting normalized image was used as the experimental input data for quantitative analysis. Multi-slice image simulations were carried out using the Dr. Probe software~\cite{barthelDrProbeSoftware2018}. The BO boosted image quantification was performed on an NVIDIA GeForce RTX 3090 graphics processing unit (GPU). Atomic models were visualized using the VESTA software~\cite{mommaVESTA3Threedimensional2011b}.

\section{Results and Discussion}

Fig.~\ref{fig: imaging parameters}(a) shows the normalized HRTEM image of BTO recorded along the [110] direction. All atomic columns (see models inserted; $\textbf{\textit x} \parallel [\bar{1}10]$ and $\textbf{\textit y} \parallel [001]$), namely BaO, Ti, and O, exhibit bright contrast under the NCSI conditions. For better demonstration, the image is partitioned into 12 sub-regions of interest (ROIs) in the crystalline area, each of which corresponds to one projected BTO [110] unit cell. The measured periodicity along [$\bar{1}$10] and [001] is 0.582 and 0.407 nm (angle: 89.9$^\circ$), respectively. According to our previous studies~\cite{geAtomicScaleObservationOffCentering2022,jinUnderstandingStructuralIncorporation2022}, an error of about $1\%$ is present in the microscope calibration. Further image quantification will be divided into three steps, aiming to retrieve the 3D atomic structure of the entire area. The three steps are discussed in the following sub-sections, including 1) quantitative image matching for averaged ROIs, 2) parallel image matching for individual ROIs, and 3) construction of an integrated 3D atomic structure and a final visualization via simulation using this integrated model.

Before proceeding, we define two sets of imaging parameters. The global parameters (indicated by subscript "global") are those applied to the final integrated model representing the entire field of view, whereas the ROI-related parameters (indicated by subscript "ROI") are determined from an individual or averaged ROI. Except for $C$1, the other imaging parameters are identical (under ideal conditions) in the two sets and therefore written without explicit subscripts for simplicity. The relationship between $C1_{\mathrm{ROI}}$  and $C1_{\mathrm{global}}$ will be addressed in Section 4.3.

\subsection{Quantitative image matching for averaged ROIs}
\label{subsec:4.1}

\begin{figure}[htbp] 
   \centering
   \includegraphics[width=\linewidth]{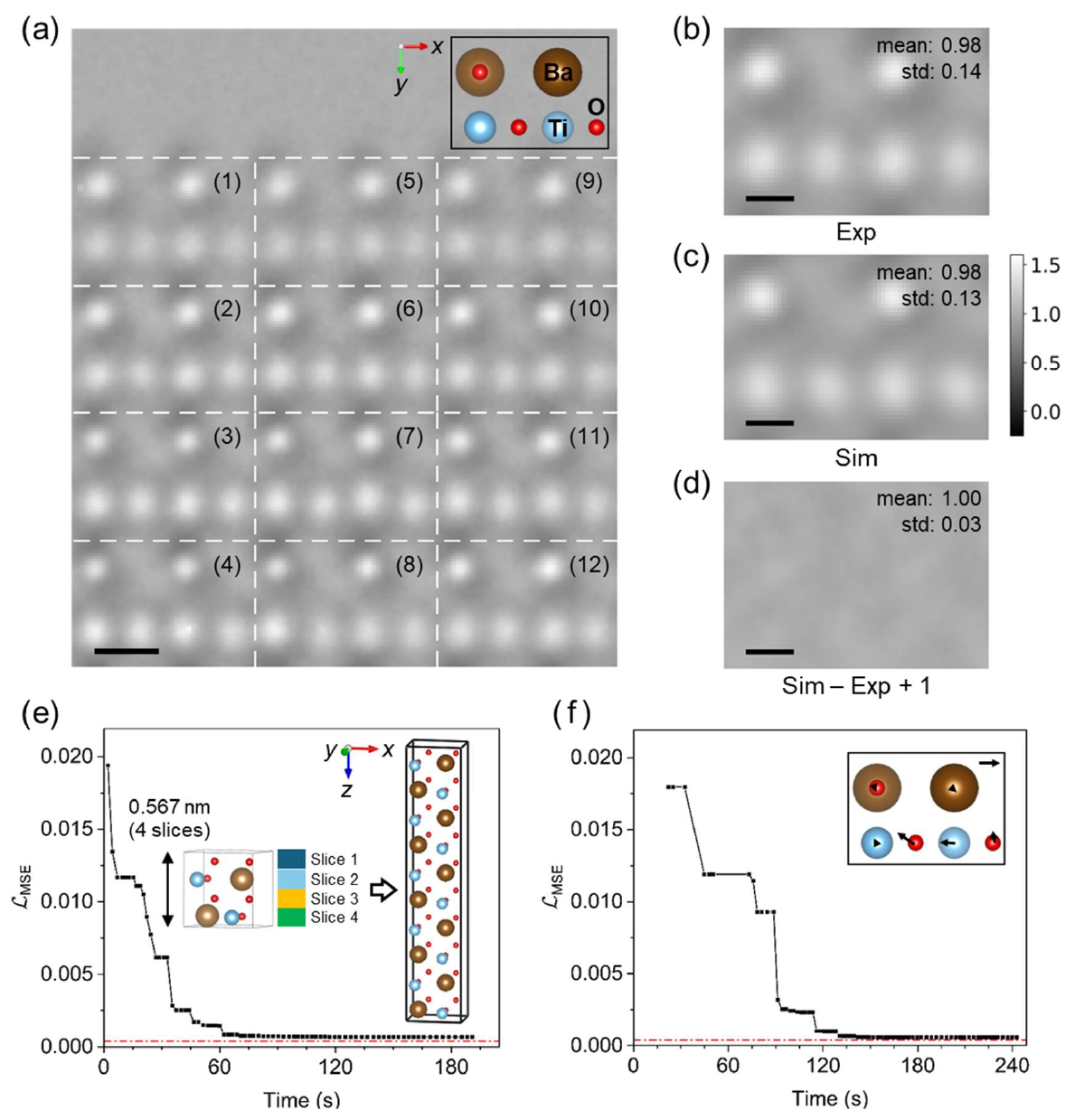}
   \caption{{\bf Determination of imaging parameters based on averaged experimental image}. (a) Normalized experimental image of BTO and (b) averaged image from ROI2, ROI6 and ROI10 in (a) used as input (\textit{i.e.}, Exp) for global imaging parameter fitting. Inset to (a) shows a cubic BTO model projected along the [110] direction. The model size outlined by the black frame corresponds to a single ROI. (c) The best-fit simulated image (\textit{i.e.}, Sim) and (d) difference image between Exp and Sim after image parameter fitting. A unity value is added to the mean intensity of (d) to allow all images to be displayed on the same absolute intensity scale (see greyscale bar alongside (c)). Scale bars: 0.2 nm in (a) and 0.1 nm in (b)-(d). (e) Minimum $\mathcal{L}_\mathrm{MSE}$ value at each iteration plotted as a function of simulation time, showing a stable convergence after 75 s for imaging parameter fitting. The inset shows that a sample thickness of 2.835 nm (\textit{i.e.}, 20 slices) is applied. (f) Minimum $\mathcal{L}_\mathrm{MSE}$ value plotted as a function of time for subsequent structure parameter fitting. Arrows in the inset denote atomic column shifts after structure parameter optimization. Scale bar: 5 pm. Red dashed lines mark $\mathcal{L}_\mathrm{MSE}$ measured from vacuum. The time preceding the first data point corresponds to initialization and scales linearly with the number of optimized parameters.}
   \label{fig: imaging parameters}
\end{figure}

In the first step, we perform quantitative matching between simulation and an averaged experimental image constructed from selected ROIs, as shown in Fig.~\ref{fig: imaging parameters}(b)-(e), with the primary task to figure out the global imaging parameters. The selected ROIs are chosen manually based on similar image characteristics, such as atomic column intensities and contrast, so that they are likely areas of comparable sample thickness. Although averaging effectively also reduces noise, its primary role is to suppress local structure fluctuations (such as variations in surface configuration) in the selected ROIs. Averaging the experimental image thus allows us to assume an effectively homogeneous underlying structure that can be processed by sequentially repeating slices prepared from a single unit-cell model (inset in Fig.~\ref{fig: imaging parameters}(e)). This enables imaging related parameters to be optimized with a substantially reduced set of structure parameters (\textit{i.e.}, without considering individual column configurations, as will be performed later in Section~\ref{subsec:4.2}). Structure parameters for the unit cell are subsequently refined, and the optimization proceeds iteratively by alternating between imaging and structure parameters until joint convergence is achieved. Preferably, when present in the same image, the averaged ROI image should be prepared from well-defined reference materials, such as Si or SrTiO$_3$ substrates, for which the structure parameters are considered as well-known. In such a case, no further iteration would be required for determining imaging parameters.

Fig.~\ref{fig: imaging parameters}(b)-(d) shows the best-fit between the simulated and the averaged experimental image patch. This fit provides a starting set of imaging parameters and an estimate of the sample thickness of the BTO model applied (20 slices = 2.835 nm, as visualized by the inset to Fig.~\ref{fig: imaging parameters}(e)). The respective parameter values are summarized in Table~\ref{tab:matching_results}. 
Here, a scaling factor of 0.98 has been applied to the experimental sampling rate to allow comparisons with a cubic BTO test model (lattice parameter a = 0.4006 nm) adopted from ICSD 67518~\cite{buttnerStructuralParametersElectron1992}. Four slices were prepared in the viewing/thickness direction (\textit{i.e.}, [110] aligned with $\textbf{\textit{z}}$). 
Fig.~\ref{fig: imaging parameters}(e) plots the $\mathcal{L}_\mathrm{MSE}$ value as a function of the iteration time for the imaging parameter matching. It is seen that a stable convergence can be reached within about 75 s boosted by BO, which is strikingly faster than that of the previous user-supervised process, usually only finished after several days to weeks. The converged $\mathcal{L}_\mathrm{MSE}$ value is about 0.0007, which is slightly higher than the noise level measured from vacuum areas (0.0004), suggesting an excellent fit.

As an alternative, the fit is also evaluated using the mean values of the image intensity and the corresponding std as figures of merit. Again, both mean and std values show excellent match (Fig.~\ref{fig: imaging parameters}(b), (c)), which results in a rather homogenous difference image, with a slightly higher std as compared with the vacuum noise level of 0.013. To display all images on the same absolute intensity scale, a unity value has been added to the difference image shown in Fig.~\ref{fig: imaging parameters}(d). Unless specified, the same operation will be used for visualization elsewhere in this work. In addition, the peak $NCC$ value ($NCC_{\rm P}$) between experiment and simulation, shown in Fig.~\ref{fig: imaging parameters}(b), (c) is 98.1\%, located at the center of the calculated $NCC$ map. Hereafter, only the $NCC_{\rm P}$ values are reported, which by default correspond to the central values of the $NCC$ maps, indicating that no image shift is detected between experiment and simulation.

We now preserve the imaging parameters and optimize the structure parameters of BTO. Since minor variation in the \textit{z} (fractional) coordinate of an atom (\textit{i.e.}, \textit{z} position changes within the same slice) can hardly cause detectable changes in the image contrast, only \textit{x} and \textit{y} coordinates are improved iteratively. For atoms lined up in the same column (\textit{e.g.}, Ba and O in the BaO column, Fig.~\ref{fig: imaging parameters}(a)), we set identical changes in \textit{x} and \textit{y} to keep columns intact and move the whole column. In addition, to consider an average vacancy concentration, we also optimize the occupancy values included in the structure model. Here, the same fractional occupancy will be applied to the same type of atoms in the same column, by effectively reducing the strength of the projected atomic scattering potential. In total, 12 position parameters (\textit{x} and \textit{y} for 6 atomic columns) and 8 occupancy parameters are thus optimized.

Results of the structure parameter optimization are visualized in Fig.~\ref{fig: imaging parameters}(f). Only subtle atomic column shifts are required ($< 5$ pm), which is lower than the measurement precision~\cite{geAtomicScaleObservationOffCentering2022}. As for occupancy, most of the values are found to be 100\%, except two O with values of 98\% and 90\%, suggesting a very low vacancy concentration in the measured area. $\mathcal{L}_\mathrm{MSE}$ shows very minor improvement, meaning that further iteration between imaging and structure parameters is not needed. Therefore, we finish Step 1 and take the imaging parameters listed in Table~\ref{tab:matching_results} as the global values for the subsequent optimizations in Step 2 (Section~\ref{subsec:4.2}).

It should be emphasized that besides the primary task, Step 1 can also be applied to the determination of unknown structures (\textit{e.g.}, phase transitions and structural relaxations). Indeed, the present BTO provides a good example where the average structure transforms from room temperature tetragonal bulk phase to a cubic-like form in thin regions (with reduced Ti and O dipolar shifts in the [001] direction), evidenced by our quantitative image matching. Further application to a YAlO$_3$ (YAO) sample \cite{jinAtomicResolutionImaging2017} is given in Fig. S1, and the best-matching parameters are listed in Table S1, showing that the BO boosted method applies successfully to different materials and experimental conditions.

\begin{table}[htbp]
  \caption{Parameters for the best-fit simulation. The azimuth of the simulation parameters is given with respect to the $\textbf{\textit{x}}$ direction in Fig.~\ref{fig: imaging parameters}. Isotropic Debye-Waller factors $ B $ for Ba, Ti, and O, and the lattice parameter $a$ were taken from~\cite{buttnerStructuralParametersElectron1992} as known parameters.}
  \centering
  \label{tab:matching_results}
  \setlength{\tabcolsep}{10pt}
  \begin{tabular}{llrr}
    \toprule
    Sub-space     & Parameter     & Magnitude     & Azimuth \\
    \midrule
                  & semi-convergence angle & 0.46 mrad & \\
                         & focus spread  & 4.00 nm & \\
                  & defocus $C1$ (for averaged ROI)    & +4.58 nm &       \\
                  & two-fold astigmatism $A1$   & 2.57 nm & $116.7^\circ$  \\
                  & three-fold astigmatism $A2$ & 70.49 nm  & $-28.0^\circ$   \\
                  & coma $B2$     & 99.48 nm & $138.2^\circ$  \\
    Imaging related & 3rd order spherical aberration $C3$    & \SI{-5.00}{\micro\meter}  &   \\
    parameter     & star aberration $S3$ & \SI{2.83}{\micro\meter}& $-45.0^\circ$ \\
                  & fourfold astigmatism $A3$ & \SI{2.83}{\micro\meter}& $-135.0^\circ$ \\
                  & 5th order spherical aberration $C5$ & +2.00 mm  &     \\
                  & image spread $\sigma$ & 15.5 pm  &   \\
                  
    \midrule
    Structure     & Specimen thickness & 2.835 nm  &       \\
    parameter     & Specimen tilt     & 5.32 mrad & $2.9^\circ$      \\  
                  & relative absorption \textit{AF}    & 1.9\% &     \\
    \midrule
    Known      
    parameter     & $B_\mathrm{Ba}$     & 0.00427 nm$^2$ &  \\
                  & $B_\mathrm{Ti}$    & 0.00695 nm$^2$ &   \\
                  & $B_\mathrm{O}$    & 0.00561 nm$^2$ &  \\
                  & Lattice parameter $a$ & 0.4006 nm & \\
    \bottomrule
  \end{tabular}
\end{table}

\subsection{Parallel matching for 3D structural determination of individual ROIs}
\label{subsec:4.2}

In this section, we continue to use BO to retrieve the 3D structure of BTO. In order to speed up the optimization process, we perform 3D structure determination in parallel for each ROI separately. The results are then stitched together to reconstruct a complete and larger 3D model of the entire area. Prior to this, optimization of \textit{x} and \textit{y} coordinates has been performed for each ROI, following the procedure described in Section~\ref{subsec:4.1}. The validity of parallel matching lies in the fact that contrast delocalization for thin samples is confined within a sub-angstrom level by aberration correction \cite{urbanProgressAtomicresolutionAberration2023} and individual ROIs can thus be treated almost independently. 

For each ROI, the three parameters listed in Table~\ref{tab:matching_results} will be optimized together with an extended parameter set describing the local 3D atomic structure model of the sample. First, optimization of the 3D atomic model unavoidably changes the thickness $t$ of each ROI (denoted as $t_{\mathrm{ROI}}$). Second, \textit{AF} describes the loss of electrons in the image forming process, and its influence on the final image depends on the local arrangement of a material, for instance, local thickness and tilt. In this sense, \textit{AF} should be practically processed as an ROI related parameter (denoted as \textit{AF}$_{\mathrm{ROI}}$). Third, the parallel matching treats ROIs as independent objects, which neglects their relative position relationships in the electron beam (thickness) direction. At this stage, the defocus value $C$1 will be optimized instead for each ROI (denoted as $C1_{\mathrm{ROI}}$). Detailed discussion will be made in Section~\ref{subsec:4.3}.

\begin{figure}[htbp] 
   \centering
   \includegraphics[width=1\columnwidth]{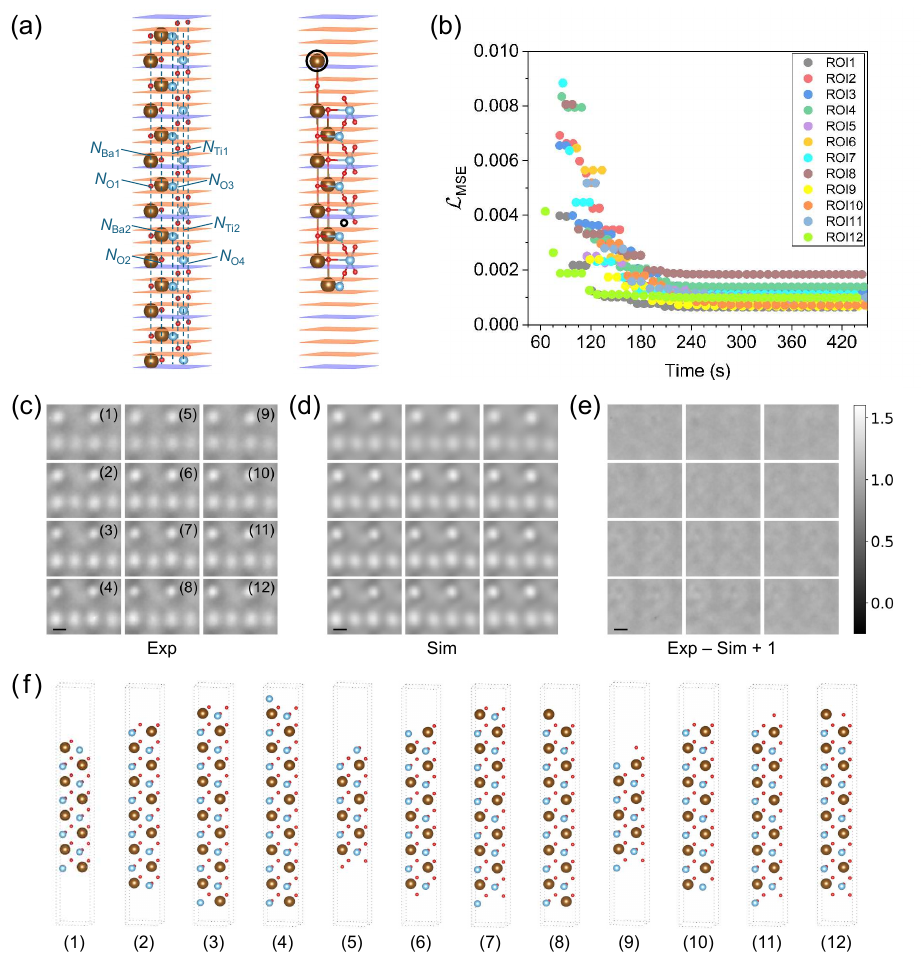} 
   \caption{{\bf Parallel matching for 3D structural determination of ROIs}. Introduced parameters $N$ (\textit{i.e.}, the number of atoms for each type in each atomic column) to optimize the 3D atomic model of each ROI. Individual vacancies inside a column (small circle) are not treated explicitly, and isolated surface atoms (large circle) are considered energetically unfavorable. (b) Plots of $L_{\rm MSE}$ as a function of iteration time for all ROIs, evidencing a steady convergence after about 300 s. (c)-(e) Experimental images, best-matching simulations, and corresponding difference images for all ROIs, respectively, displayed at the same greyscale level. (f) Determined 3D atomic models for all ROIs, showing the distinct increase of sample thickness from top to bottom (wedge shape), resulting from TEM sample preparation.}
   \label{fig: 1}
\end{figure}

To optimize the 3D atomic model, additionally, 8 (discrete) parameters are introduced, which are the number of atoms for each type in each atomic column (\textit{i.e.}, numbers of O in two BaO and two O columns, numbers of Ba in two BaO columns, and numbers of Ti in two Ti columns). Since in Section~\ref{subsec:4.1} we have already obtained an averaged thickness (\textit{i.e.}, 20 slices, Fig.~\ref{fig: imaging parameters}(e)), we use this value to help define the maximum thickness (here 28 slices, Fig.~\ref{fig: 1}(a)), so that the range of each introduced variable is limited. We also make the following assumptions: 1) Within each atomic column, atoms are supposed to stack continuously. Therefore, individual vacancies inside a column, such as the O vacancy marked by the small circle in Fig.~\ref{fig: 1}(a), are not treated explicitly. Instead, their effects are incorporated in the average fractional occupancy of O sites in a column, as already used in Section~\ref{subsec:4.1}. 2) Isolated surface atoms that maintain unreasonable $x$-$y$ distances from the surrounding atomic environment, \textit{e.g.}, the Ba atom marked by the large circle in Fig. 3a, are rejected. In this case, the presence of only a single atom across two consecutive slices is considered energetically unfavorable.

Fig.~\ref{fig: 1}(b) plots the $\mathcal{L}_\mathrm{MSE}$ value as a function of iteration time for all ROIs. It is seen that a steady convergence can be reached after about 300 s, which is again dramatically faster than user-supervised reconstructions often taking weeks or months. Fig.~\ref{fig: 1}(c)-(e) shows the quantitative matching results for all 12 ROIs. The three parameters optimized are listed in Table~\ref{tab:region_metrics} together with the evaluation statistics, and the 3D atomic models for each ROI are shown in Fig.~\ref{fig: 1}(f). Distinct increase of sample thickness is evidenced from top to bottom ROIs as shown in Fig.~\ref{fig: 1}(c). This wedge-shaped thickness variation is a result of TEM sample preparation.

From Table~\ref{tab:region_metrics}, the fit in Fig.~\ref{fig: 1}(c)-(e) appears to be quantitatively good, although slightly higher std values is observed in the difference image of some ROIs (\textit{e.g.}, ROI4 and ROI8). Since the imaging parameters in Section~\ref{subsec:4.1} are optimized for the averaged image of ROI2, ROI6 and ROI10, better fitting can be naturally expected for ROIs nearby. In particular, the bottom ROIs have increased thickness, which may require deviating imaging parameters from those determined in Table~\ref{tab:matching_results}. In this regard, a combined matching by using multiple averaged experimental images from ROIs with different thicknesses could be a solution to determine better global imaging parameters for the entire area.

\begin{table}[htbp]
  \centering
  \caption{Simulation results for 12 regions}
  \label{tab:region_metrics}
  \setlength{\tabcolsep}{4pt}
  \small
  \begin{tabular}{cccccccccccc}
    \toprule
    ROI &
    \makecell{$t_{\mathrm{ROI}}$} &
    \makecell{\textit{AF$_{\mathrm{ROI}}$}} &
    \makecell{C1} &
    \multicolumn{2}{c}{Exp} &
    \multicolumn{2}{c}{Sim} &
    \multicolumn{2}{c}{Sim--Exp} &
    \makecell{$\mathcal{L}_\mathrm{MSE}$} & 
    \makecell{\textit{NCC}$_{\rm P}$\\} \\
    \cmidrule(lr){5-6}
    \cmidrule(lr){7-8}
    \cmidrule(lr){9-10}
    & (nm)& (\%)& (nm)&
    mean & std &
    mean & std &
    mean & std &(/100)
    &(\%) \\
    \midrule
    1  & 2.27 & 1.8 & +4.31 & 0.99 & 0.12 & 0.99 & 0.11 & 0.00 & 0.03& 0.07 & 97.5 \\
    2  & 2.83 & 1.7 & +4.41 & 0.98 & 0.14 & 0.98 & 0.13 & 0.00 & 0.03 & 0.08 & 97.8 \\
    3  & 3.40 & 2.5 & +4.34 & 0.97 & 0.14 & 0.97 & 0.14 & 0.00 & 0.04 & 0.12 & 97.0 \\
    4  & 3.54 & 2.7 & +4.28 & 0.97 & 0.15 & 0.96 & 0.14 & 0.00 & 0.04 & 0.14 & 96.8 \\
    5  & 2.12 & 2.2 & +4.12 & 0.99 & 0.11 & 0.99 & 0.10 & 0.00 & 0.03 & 0.08 & 96.6 \\
    6  & 2.97 & 2.0 & +4.63 & 0.98 & 0.14 & 0.98 & 0.13 & 0.00 & 0.03 & 0.08 & 98.0 \\
    7  & 3.40 & 2.4 & +4.27 & 0.97 & 0.15 & 0.97 & 0.14 & 0.00 & 0.03 & 0.11 & 97.3 \\
    8  & 3.26 & 2.5 & +4.60 & 0.97 & 0.14 & 0.97 & 0.13 & 0.00 & 0.04 & 0.19 & 95.4 \\
    9  & 2.12 & 2.5 & +4.17 & 0.99 & 0.10 & 0.99 & 0.10 & 0.00 & 0.03 & 0.07 & 96.7 \\
    10 & 2.83 & 2.1 & +4.53 & 0.98 & 0.13 & 0.98 & 0.13 & 0.00 & 0.03 & 0.07 & 98.0 \\
    11 & 3.26 & 2.0 & +4.28 & 0.98 & 0.14 & 0.98 & 0.14 & 0.00 & 0.03 & 0.10 & 97.4 \\
    12 & 3.11 & 1.6 & +4.75 & 0.98 & 0.15 & 0.98 & 0.14 & 0.00 & 0.03 & 0.10 & 97.8 \\
    \bottomrule
  \end{tabular}
\end{table}

\subsection{Building 3D atomic structures and performing the corresponding simulation}
\label{subsec:4.3}

\begin{figure}[htbp] 
   \centering
   \includegraphics[width=1\columnwidth]{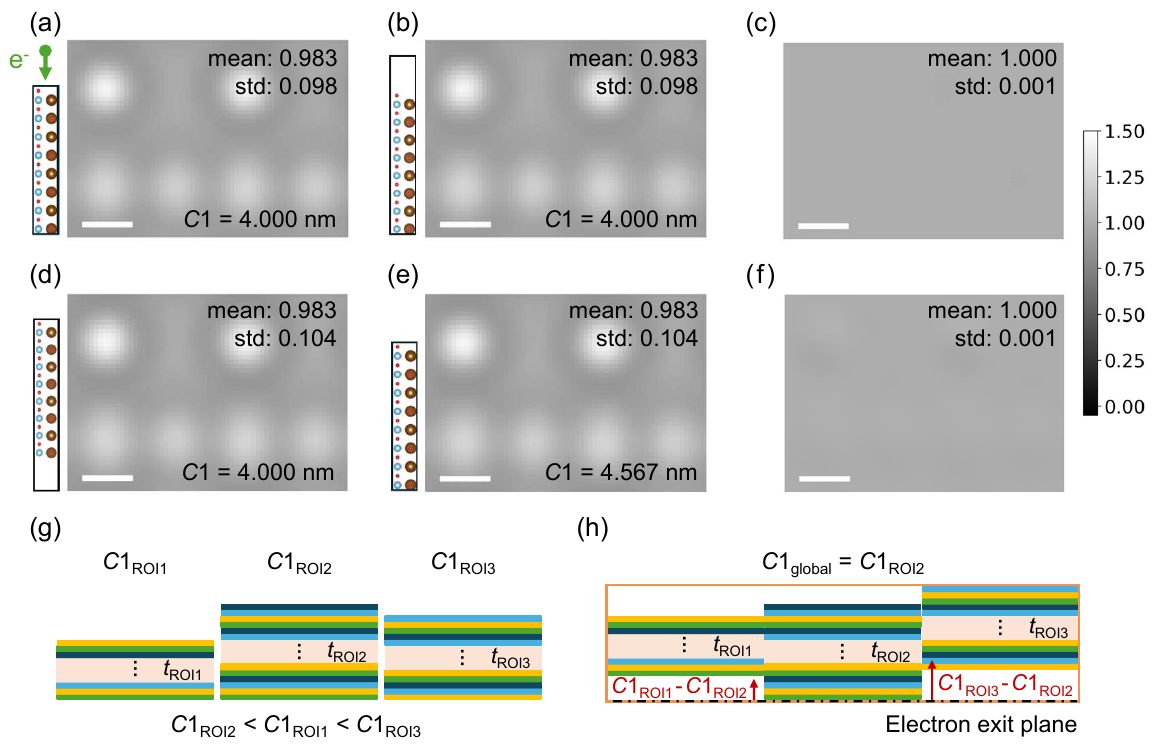} 
   \caption{{\bf Determination of the global $C1$ and relative position relationship between ROIs}. (a) and (b) Simulated images for two BTO supercell models under identical simulation parameters. The model in (b) consists of the same crystalline part of (a) plus a vacuum with thickness of 0.567 nm, equivalent to 4 slices of BTO. (c) Difference image between (a) and (b) showing a negligible contribution of added vacuum at the electron entrance plane. (d) Simulated image for a BTO model, now with a vacuum added below the sample exit plane, while other parameters are maintained the same. A change in the image contrast (std) is detected. (e) Simulated image for the same BTO model as in (a), but with an addition to $C1$ of 0.567 nm, compensating the vacuum volume added below the sample in (d). (f) Difference image between (d) and (e) showing their equivalence. (g) and (h) illustrate how structure models of separately evaluated patches can be combined into one model by interpreting local defocus differences as relative z-position differences with respect to a common reference plane. Scale bar: 0.1 nm.}
   \label{fig: thichnessad}
\end{figure}

As mentioned in Section~\ref{subsec:4.2}, the relative position relationship between different ROIs in the thickness direction has not been considered. Instead, we applied $C$1 as a ROI related parameter, which appears to be contradictory to the consensus that a global $C$1 (denoted as $C1_{\mathrm{global}}$) should be used for the entire area during HRTEM imaging. In this section, we first demonstrate that the determined $C$1$_{\mathrm{ROI}}$ can be used to figure out $C$1$_{\mathrm{global}}$ and the relative position relationship between ROIs along the incident beam direction $\textbf{\textit{z}}$, as shown in Fig.~\ref{fig: thichnessad}.

Fig.~\ref{fig: thichnessad}(a), (b) show the simulation results for two supercell models of BTO after applying identical simulation parameters (only $C$1 is specified). The model in Fig.~\ref{fig: thichnessad}(b) consists of the same crystalline part shown in Fig.~\ref{fig: thichnessad}(a) plus a vacuum located at the electron entrance plane. Without loss of generality, here the thickness of vacuum is set to 0.567 nm, equivalent to 4 slices of BTO (= 1 minimum repeating unit in the thickness direction). It is evidenced from the intensity statistics and the difference image in Fig.~\ref{fig: thichnessad}(c) that these vacant slices have no physical contribution to the image contrast. In contrast, when the same vacuum is added below the exit plane, as in Fig.~\ref{fig: thichnessad}(d), an obvious change in the std is observed. This change of contrast can be reproduced by applying a larger $C$1 value to the case of Fig.~\ref{fig: thichnessad}(a) (\textit{i.e.}, an increased defocus of 0.567 nm, see Fig.~\ref{fig: thichnessad}(e)), as confirmed by the intensity statistics shown in Fig.~\ref{fig: thichnessad}(d)-(f). This means that the contribution due to the local difference of $C$1 can be converted into the effect of adding vacuum (with thickness equal to the $C$1 difference value) to the exit plane. This offers a practical solution for determining relative local z-positions of the sample exit surface while registering a global C1 value relative to a common plane below the combined sample, as illustrated in Fig.~\ref{fig: thichnessad}(g), (h).

For individual ROIs with different $C$1$_{\mathrm{ROI}}$ (\textit{e.g.}, $C$1$_{\mathrm{ROI2}}$ $<$ $C$1$_\mathrm{{ROI1}}$ $<$ $C$1$_\mathrm{{ROI3}}$, Fig.\ref{fig: thichnessad}(g)), the $C$1$_{\mathrm{global}}$ can be set to the minimum value of $C$1$_{\mathrm{ROI}}$(\textit{i.e.},$C$1$_{\mathrm{global}}$ = $C$1$_{\mathrm{ROI2}}$ in Fig.~\ref{fig: thichnessad}(g)). To compensate for the defocus induced contrast change, ROI1 (or ROI3) will be shifted away from the exit plane of ROI2 by a distance of $C$1$_{\mathrm{ROI1}}$ – $C$1$_{\mathrm{ROI2}}$ (or $C$1$_{\mathrm{ROI3}}$ – $C$1$_{\mathrm{ROI2}}$), as indicated by red arrows. In practice, this shift value will first be calculated in units of slice thickness and then rounded to the neighboring slice (with identical structure) to preserve the continuity of the planes (if no antiphase defect is experimentally confirmed). With this operation, an integrated supercell model can be constructed, as outlined by the orange frame in Fig.~\ref{fig: thichnessad}(h).

\begin{figure}[htbp] 
   \centering
   \includegraphics[width=1\columnwidth]{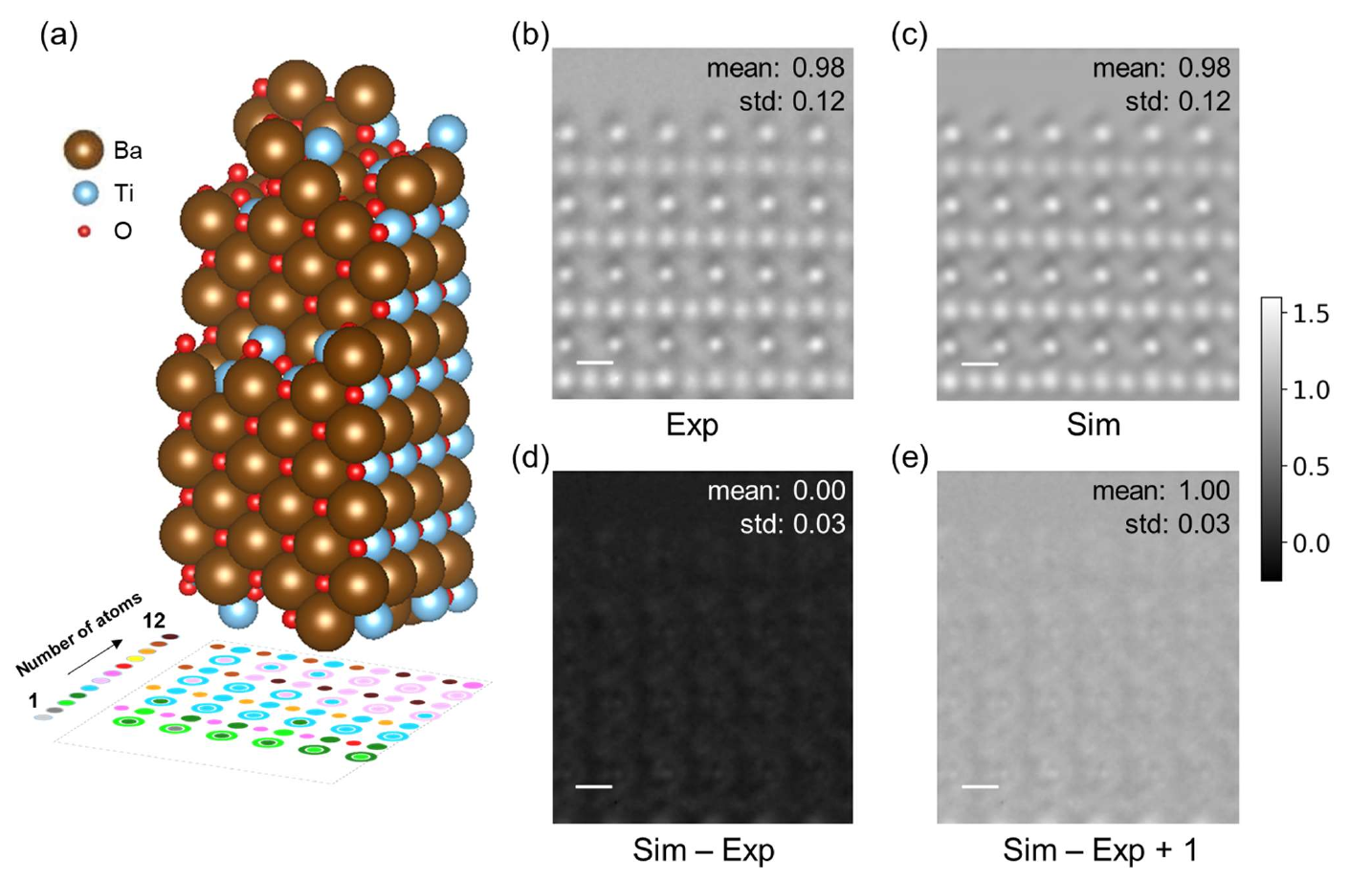} 
   \caption{{\bf Simulation using the integrated 3D atomic structure}. (a) Integrated supercell model constructed from all the individual models. Colored circles represent the number of atoms in atomic columns. Large: Ba; Medium: Ti; Small: O. (b) Reproduction of the experimentally recorded image shown in Fig.~\ref{fig: imaging parameters}(a) and (c) the simulated image using the integrated 3D atomic model for the entire area. The $NCC_{\rm P}$ between (b) and (c) is 96.3\%. (d) and (e) Original and mean-intensity-adjusted difference images between (b) and (c), showing the residual image contrast. Scale bars: 0.2 nm. All images are displayed at the same grey-scale level.}
   \label{fig: whole sample}
\end{figure}

Now we construct an integrated 3D atomic model (Fig.~\ref{fig: whole sample}(a)) by stitching all the individual models shown in Fig.~\ref{fig: 1}(f) and perform a simulation for the entire area shown in Fig.~\ref{fig: imaging parameters}(a) using the integrated model. For this simulation, a compromise on the electron absorption factor $AF$ has to be made to reach a global value for the simulation model. The results are shown in Fig.~\ref{fig: whole sample}(b)-(e). Here, Fig.~\ref{fig: whole sample}(b) is a reproduction of Fig.~\ref{fig: imaging parameters}(a), and Fig.~\ref{fig: whole sample}(c) is the image simulated using the integrated model. The $C$1$_{\mathrm{global}}$ used is +4.12 nm (= $C$1$_{\mathrm{ROI5}}$, see Table~\ref{tab:region_metrics}) and the \textit{AF}$_{\mathrm{global}}$ is 2.2\% (averaged from values of all ROIs, see Table~\ref{tab:region_metrics}), other parameters remain unchanged as listed in Table~\ref{tab:matching_results}. Fig.~\ref{fig: whole sample}(d), (e) show the original and the mean-intensity-adjusted difference images, displayed at the same greyscale level as in Fig.~\ref{fig: whole sample}(b), (c), revealing the residual image contrast: the original difference preserves the signed residuals, whereas the adjusted image enhances the visibility of subtle spatial variations. From the image statistics, a good fit between experiment and simulation can be concluded, suggesting the validity of the proposed solution.

It should be noted that several factors can influence the std of the difference image (Fig.~\ref{fig: whole sample}(d), (e)). First, \textit{AF}\textsubscript{ROI} is employed in the parallel matching process. In contrast, only a global value (\textit{AF}\textsubscript{global} = 2.2\%) can be applied in the simulation software package, which introduces extra intensity fluctuations in the comparison over the entire area. For large experimental image areas, particularly those exhibiting pronounced variations in thickness and tilt and/or comprising different materials (thus distinct \textit{AF}\textsubscript{ROI}), simulation based on an integrated supercell and an \textit{AF}\textsubscript{global} (though local thickness, tilt, and \textit{AF} can be measured by parallel ROI matching) may lead to pronounced difference with respect to the experiment. A more refined treatment of \textit{AF} within the physics-informed model is therefore desirable.  Second, when constructing the integrated supercell model, as shown in Fig.~\ref{fig: thichnessad}(h), the maximum rounded shift applied to each ROI can reach half of the repeating unit. For BTO, this corresponds to 2 slice thicknesses ($\approx$0.28~nm, see Fig.~\ref{fig: imaging parameters}(e)). Residual errors associated with these rounded $z$-shifts consequently increase the intensity fluctuations. In addition, the combined $x$-$y$ shifts of all atoms in a column, together with the assumption of average fractional site occupancies, also contribute to increased intensity differences.

\subsection{Discussion}
\label{subsec:4.4}

By the procedure described above, the 3D atomic structure of a nanometer sized BTO has been successfully determined with an automated optimization loop and significantly reduced human effort compared to previous approaches. All parameters used in the BO-boosted quantitative HRTEM are summarized in Table S2, in which 17 unknown parameters (corresponding to 48 individual variables) are eventually fitted. Instead of simultaneously fitting all parameters~\cite{jiaDetermination3DShape2014}, the present approach optimizes parameters through three basic procedures organized into two steps: Step 1 comprises (i) optimization of imaging related parameters and (ii) (average) structural matching, while Step 2 involves (iii) 3D structural determination. The parameters optimized in each procedure are listed in Table S3, together with other details on the acquisition function, the surrogate model and the time consumption. It is seen that the number of variables to be optimized in each procedure is about 10-20, which is well within TuRBO’s capabilities. Although it is theoretically feasible to perform simultaneous optimization of all parameters over a larger specimen area (given that $\textit{AF}_{\rm ROI}$ variations are relatively subtle), such an approach in practice would incur excessively high computational costs and require GPUs with substantial memory capacity. In particular, the number of parameters in the 3D structural determination scales linearly with the size of specimen area (\textit{i.e.}, number of ROIs). More efficient sampling strategies, \textit{e.g.}, using variational autoencoder~\cite{slautinIntegratingPredictiveGenerative2025,biswasOptimizingTrainingTrajectories2023} or ensemble-based recommendations\cite{shenSuperSaltEquivariantNeural2025}, can be applied to tackle optimization in higher ($\sim 10^2$ – $10^3$) dimensional spaces. In this regard, our parallel matching thus provides an effective means of mitigating the challenges associated with high dimensionality.

It is also apparent that the probabilistic-based algorithm shows its superiority over the conventional user-supervised iteration method, thus greatly improving the quantification efficiency. With ROI segmentation and parallel matching, the current approach allows the 3D structure to be determined within a few minutes. This makes the method particularly suited for automated experimental scenarios, opening up new possibilities for future studies on larger specimen volumes, particularly those containing heterogeneous structures. Moreover, the substantial improvement in optimization efficiency achieved by the present BO framework also suggests its potential extension toward quantitative HRTEM analysis in the time domain. Here, we do not refer to extremely high frame-rate imaging (\textit{e.g.}, hundreds of frames per second or above). Instead, at the current stage, our approach is particularly suited for tracking relatively sluggish structural evolutions that occur on the time scale of seconds to minutes, such as defect-mediated phase transitions, domain wall migration, surface reconstruction, and oxygen-vacancy redistribution under moderate stimuli~\cite{xingAtomicscaleOperandoObservation2022,jeongSubsurfaceOxygenVacancy2025,weiUnconventionalTransientPhase2020,weiSituObservationPointDefectInduced2021}. We therefore believe that this framework can serve as an enabling tool for \textit{in situ/operando} HRTEM studies of structural evolution processes that are slow to be resolved at conventional acquisition rates, yet complex enough to require rigorous model-based quantification.

Our quantitative BO-boosted HRTEM analysis reveals a clear reduction of the Ti–O relative displacement in the edge region of BTO. Because the Ti off-center shift with respect to the surrounding O octahedron directly represents the structural order parameter of tetragonal BTO, the observed decrease indicates a local suppression of ferroelectric distortion and a tendency toward a more cubic-like average configuration. The systematic nature of this trend across multiple ROIs, together with the robustness of the BO-matching procedure, suggests that the displacement reduction is intrinsic rather than an artifact of imaging imperfection or fitting uncertainty. This behavior can be understood from the following perspectives: First, bulk BTO undergoes a tetragonal-to-cubic transition at ~120 °C~\cite{harwoodCuriePointBarium1947}. In reduced dimensions, finite-size effects and incomplete screening of bound surface charges lead to an enhanced depolarization field that opposes the spontaneous polarization. First-principles calculations have demonstrated that below a critical thickness, the depolarization field destabilizes the ferroelectric state, effectively reducing both the equilibrium polarization amplitude and the Curie temperature~\cite{junqueraCriticalThicknessFerroelectricity2003}. In the edge region of our specimen, where the local thickness decreases substantially, the electrostatic boundary condition approaches a weak-screening or open-circuit limit. Under such circumstances, the free-energy gain associated with Ti off-centering is diminished, favoring a configuration closer to cubic symmetry. The reduced Ti–O displacement extracted from the BO-boosted structural refinement is therefore consistent with a depolarization-driven suppression of the ferroelectric order parameter. Second, electron-beam exposure may further perturb the local ferroelectric stability. Electron irradiation in insulating oxides is known to induce localized charging and internal electric fields, and in some cases modest temperature increases, all of which can influence domain configurations and polarization states~\cite{hartElectronbeaminducedFerroelectricDomain2016, jiangBeamDamageInduced2016}. While the absolute temperature rise under typical TEM conditions is generally limited, ultrathin regions that are already close to the stability boundary between tetragonal and cubic phases may be particularly sensitive to such perturbations. Consequently, the combined effect of reduced thickness and beam-induced electrostatic disturbance may therefore shift the local free-energy landscape toward a weakened polar distortion, contributing to the experimentally observed decrease of the Ti–O displacement.

A related electrostatic argument applies to YAO, although the driving mechanism may differ in origin. YAO consists of alternating (YO)$^{+}$ and ($\rm{AlO}_2$)$^{-}$ layers along the viewing direction~\cite{jinAtomicResolutionImaging2017}. In ultrathin geometries, such charged-layer stacking would generate a diverging electrostatic potential unless compensated, as described for polar ionic surfaces~\cite{taskerStabilityIonicCrystal1979, nogueraPolarOxideSurfaces2000}. Structural relaxation—including ionic rumpling and interlayer spacing adjustment—provides an efficient mechanism to reduce the electrostatic energy and restore charge neutrality. Given the extreme thinness of the YAO specimen examined, such relaxation is expected to be amplified. The picometer-scale column separations quantified by BO-boosted HRTEM therefore reflect not only static lattice geometry but also the structural response required for electrostatic stabilization.

\section{Conclusion}

In conclusion, we have developed a physics-informed BO framework for 3D structure reconstruction from a single HRTEM image that integrates imaging related parameter matching, atomic position optimization, and 3D structure determination. By adopting a stepwise optimization strategy and incorporating physical constraints in BO, our approach efficiently explores the high-dimensional parameter space while ensuring both computational efficiency and physical consistency. We use experimental HRTEM images of BTO and YAO as test cases and demonstrate that the algorithm can effectively reconstruct the 3D structures of specimens with good robustness. In contrast to conventional user-supervised iterative methods, our method optimizes multiple variables simultaneously via probabilistic sampling, reducing computation time from weeks or even months to minutes—corresponding to a speedup of three to four orders of magnitude. The significant improvement in efficiency not only shortens the time required to extract atomic-scale details from larger HRTEM image areas, but also advances structural quantification into the time domain, thereby laying the foundation for fully automated HRTEM workflows. Moreover, the algorithm enables rapid reconstruction of the specimen’s 3D atomic structure, offering an efficient pathway for establishing digital twins.

\section*{Acknowledgment}
The authors gratefully acknowledge the computing time provided on the high-performance computer Lichtenberg at the NHR Centers NHR4CES at TU Darmstadt. X. T. thanks the financial support from the China Scholarship Council. R.E. D.-B. and H. Z. acknowledge the funding by the Deutsche Forschungsgemeinschaft (DFG, German Research Foundation) - Project-ID 405553726 - TRR 270. This project has also received funding from the European Union’s Horizon Europe Research and Innovation Programme (Grant No. 101094299, project “IMPRESS”).

\section*{Data and Code Availability Statements}

The codes are available at the repository on GitHub (\url{https://github.com/XiankangTang/BO_HRTEM}) and \url{http://github.com/ju-bar/drprobe_clt}(Dr. Probe)). 



\bibliographystyle{IEEEtran}

\end{document}